# Enabling Synergy: Improving the Information Infrastructure for Planetary Science

Whitepaper for the Planetary Science and Astrobiology Decadal Survey 2023-2032

**Michael J. Kurtz, Alberto Accomazzi, Edwin A. Henneken**
{mkurtz,aaccomazzi,ehenneken}@cfa.harvard.edu
Center for Astrophysics | Harvard & Smithsonian
60 Garden St, Cambridge, MA 02138 USA

**Co-signers:**
Peter K. G. Williams (Center for Astrophysics | Harvard & Smithsonian; American Astronomical Society);
Mark A. Gurwell (Center for Astrophysics | Harvard & Smithsonian);
JJ Kavelaars (Canadian Astronomy Data Centre | National Research Council of Canada);
Heidi B. Hammel (AURA);
Joshua Peek (Space Telescope Science Institute);
Jonathan Fortney (University of California, Santa Cruz);
Megan E. Schwamb (Queen's University Belfast);
Bryan J. Holler (Space Telescope Science Institute);
Alice Allen (Astrophysics Source Code Library; Astronomy Department, University of Maryland College Park);
Baptiste Cecconi (Observatoire de Paris);
David Ciardi (NASA Exoplanet Science Institute - Caltech/IPAC)

## Abstract

In this whitepaper we advocate that the Planetary Science (PS) community build a discipline-specific digital library, in collaboration with the existing astronomy digital library, ADS. We suggest that the PS data archives increase their level of curation to allow for direct linking between the archival data and the derived journal articles. And we suggest that a new component of the PS information infrastructure be created to collate and curate information on features and objects in our solar system, beginning with the USGS/IAU Gazetteer of Planetary Nomenclature.

# Introduction

The development of scientific knowledge is driven by two complementary long term trends: increased specificity of research and increased diversity in the distribution of research. The number of scholarly publications we have about our natural world has been doubling every 15 years for centuries. As the fraction of all knowledge, or of a field, or a sub-field which can be mastered by any individual human decreases, the expertise of any single person becomes increasingly specific.

Complementing this trend is the long term development of communication technologies, from the printing press and libraries to electronic communication and digital storage. While individual investigations are increasingly detailed, their distribution has become increasingly diverse. The synergy arising from the interaction of these two trends is the force behind the exponential increase of scientific knowledge, and thus drives modern civilization.

Tremendous complexity is now built into our communication/information systems; digital databases and sophisticated search engines are now the norm. At the scientific level, the degree of field-specific knowledge required to communicate research is substantial, and has led to discipline-specific solutions.  Examples of data systems are the [Inorganic Crystal Structure Database](#) (ICSD), [Genbank](#), the [Protein Data Bank](#) (PDB), the [Mikulski Archive for Space Telescope](#) (MAST), and the nodes of the [Planetary Data System](#) (PDS). Digital libraries/search engines would include [PubMed](#) (medicine), [InspireHEP](#) (high energy physics), the [NASA Astrophysics Data System](#) (ADS, astronomy), [MathSciNet](#) (mathematics), and [SciFinder](#) (chemistry).

In addition, many disciplines have groups which maintain knowledge-bases providing comprehensive summaries of curated data collections. Examples of such services are [NED](#) (astronomy), the [Particle Data Group](#) (PDG, high energy physics), [Reaxys](#) (FKA: Bellstein, chemistry), and numerous others, including many commercial ventures.

In this whitepaper we advocate that the Planetary Science (PS) community build a discipline-specific digital library, in collaboration with the existing astronomy digital library, ADS.  We suggest that the PS data archives increase their level of curation to allow for direct linking between the archival data and the derived journal articles. And we suggest that a new component of the PS information infrastructure be created to collate and curate information on features and objects in our solar system, beginning with the [USGS/IAU Gazetteer of Planetary Nomenclature](#).

We also note that these developments would be of enormous benefit to the study of exoplanets and non-terrestrial biology.

# Enhancing Information Discovery

Information transfer is one of the pillars of scientific research, and the production of scholarly literature constitutes the core of this transfer. As both producers and consumers of scholarly papers, scientists are increasingly challenged by the growth and specialization of the research literature. Being able to account for the sheer increase in relevant papers is just one daunting task researchers have to contend with. There is another, equally challenging aspect of scholarly publishing: increasing complexity. Both scholarly publishing systems and the modern research life cycle itself produce a plethora of non-traditional digital objects, each of which represents particular aspects of the research being done, such as proposals, presentations, software or data products. All these digital components are interconnected and form a web of knowledge that is increasingly distributed across digital platforms and data providers.

## Publications

One of the resources used most frequently by researchers in astronomy is the ADS, a digital library and search engine which indexes the literature in Astronomy and Astrophysics. The ADS has a unique role within the NASA Astrophysics Archives in that it focuses on the scientific literature to help scientists navigate research topics and explore their connections. As interdisciplinary research develops, research fields become organically connected and discoverable through common topics, citations, and readership. By further connecting the literature with data and software products, the ADS increases discoverability of both and promotes their re-use.

The ADS is a disciplinary search system, focusing on fully representing the research literature in Astronomy & Astrophysics. Kurtz et al (2018) describe the basic intellectual structure of the ADS using a bullseye model, which we reproduce in figure 1. The three rings in this diagram represent different levels of curation applied to the bibliographic content in ADS. Levels of curation and completeness are different in each ring. The inner region, representing the Astrophysics core collection, has the highest level of curation; this segment of the ADS holdings represents that part of scholarly publications where the ADS is considered to be authoritative. In this collection, our users can expect the ADS to be complete, coverage- and citation-wise. Most of our curation efforts for this collection go into maintaining a high level of accuracy, quality and completeness, ranging from the main refereed literature to conference proceedings, theses, gray literature, software, and links to data products.

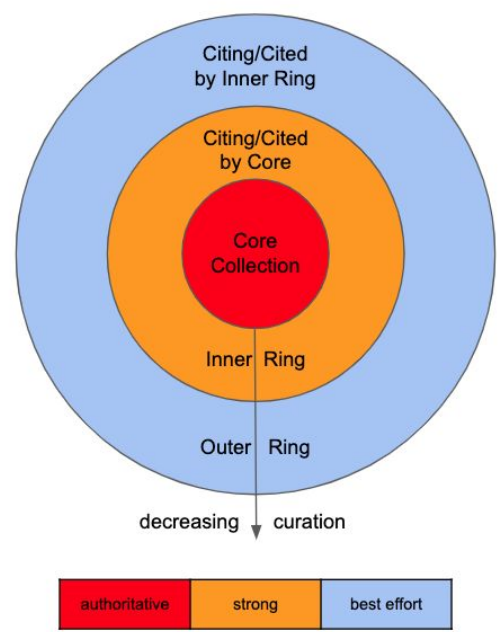

Fig 1. Curation levels of the ADS collections.

We also nurture all important collaborations necessary for this core collection.

In the next layer (inner donut) we find disciplines closely related to Astrophysics such as Planetary Sciences, High Energy Physics, and Geophysics. Here coverage is still strong, but we don't apply the same levels of curation. We still strive for completeness in the main refereed literature and the main conference proceedings series (like SPIE and AIP), but we do not seek the same kind of close collaborations as we do for Astrophysics. The outer ring are documents which might be used/cited by authors of documents in the inner ring. Here, we take publications that can be harvested easily. Curation at this level is minimal, mostly consisting of error checking.

Adding Planetary Science and Astrobiology to the Core Collection has far-reaching, trickle-down effects: additional journals, conference series, PhD Theses, preprints and technical reports will need to be identified and added to the database, their full-text mined and references indexed to enhance the citation network and search capabilities. Publications citing the newly added content will need to be identified, harvested, and their metadata ingested in the system. This includes not only traditional publications but also research objects such as cited software and data products.

To illustrate the difference between Core and Inner Ring curation we take the recent review by Kite: 2019SSRv..215...10K.  Looking at the extensive reference list we find 408 references, but ADS only has 367 (90%; for many PS articles ADS does significantly worse, some papers in e.g. astrobiology, pre-biotic chemistry, human space travel, or spacecraft engineering can be as low as 50%).  A detailed examination of the 41 missing references shows that they are a mixture of sources.  A few are from journals ADS does not cover (Sedimentology, Geo Soc Spec Pub), and books, book chapters, and conferences which ADS does not cover.  Some are from publications which ADS does have, but where the reference linking procedure did not successfully identify the citation.

All (or nearly all) of these omissions and errors would not have occurred were 2019SSRv..215...10K a core publication.  Curators would have already arranged to include the referenced journals, as well as the books and conference papers, as is routinely done for the astrophysics core.  Incomplete reference parsing problems are routinely found and fixed for core publications, but with ADS processing more than 600,000 articles per year, manual intervention is limited to core publications.

The restriction on curating non-core publications can also be seen in the record 2017LPICo2014.3047H, which is referenced by 2019SSRv..215...10K.  The conference paper has references, but they are not in a format ADS could automatically extract and recognize. Were it a core publication, that task would have been accomplished. Additionally, not a single paper in the 2019SSRv..215...10K reference list has a data link in ADS, although many of the papers use observational data, several from the Mars Curiosity Rover. A similar list of papers in the astrophysics core would have yielded over 100 data links, a result of close collaborations between the archives, data centers, journals, and the ADS.

Adding content to the ADS is but the first step required to make research content accessible to an end user. Discoverability is significantly improved by curation and the use of discipline-specific semantics, which help identify the different ways in which concepts are expressed and linked to each other in the research literature.  As an example, the dwarf planet Eris's original designation by the MPC was "2003 UB313." It is expected that a proper retrieval system be capable of finding all papers mentioning this object irrespective of the nomenclature used in them and that these papers be connected to articles sharing similar concepts drawn from the Unified Astronomy Thesaurus.

## Data and Software

We are now seeing an evolution in the way a variety of non-traditional content is considered not only relevant but rather essential to the research process. Transparency, reproducibility, and credit are the three major reasons behind current publishing trends which encourage data and software citation. While not traditional articles, all of these resources have begun appearing in citation lists of papers published in the Physical Sciences, and as the Space and Earth Science communities are moving to support data and software citations, it becomes crucial for the research infrastructure to support these efforts.

Recent surveys of Earth and Planetary scientists (Tenopir et al 2020) have found that the research community is now fully behind the principle of Data Sharing, a pillar of the Open Science Initiative and a requirement of FAIR (Findable, Accessible, Interoperable, Reusable) data principles (Wilkinson et al 2017). However, the same survey finds that the implementation of data sharing practices is still plagued by the failure to follow best practices and a lack of general support for discovering FAIR data. Raw and high-level data, when shared, end up on a variety of disparate platforms which include discipline-based archives, publisher-managed repositories, institutional repositories, or other general purpose archives. With this level of fragmentation, one major impediment to data reuse becomes discoverability.

Community efforts in promoting data sharing, such as the one sponsored by the Coalition for Publishing Data in the Earth and Space Sciences and ESIP, have led to the implementation of data citation guidelines and the implementation of data citation requirements by major publishers such as the AGU. Data exchange protocols and standards have led to the creation of technical solutions such as SCHOLIX to support the aggregation and exchange of links between literature and data products. These efforts go a long way in creating an environment in which open science can take place, but need to be complemented by an equivalent set of curatorial initiatives which bring the relevant resources back to the research communities.

The ADS has for a long time included in its database non-traditional scholarly resources such as data catalogs, observing proposals, research software packages, and technical reports and standards. Two main reasons for including these scholarly artifacts in a system such as the ADS is the need to make them easily discoverable and citable, two essential steps in making them compliant with the FAIR data principles and in support of broader Open Science goals. The ADS, in collaboration with the AAS and Zenodo, has already implemented a workflow which allows it to detect and ingest citations to software products used in scholarly publications (Muench et al 2020). This has in turn further enabled the citation of software as a first-class research object and has increased the adoption of software citation in articles indexed in the ADS (see Figure 2).

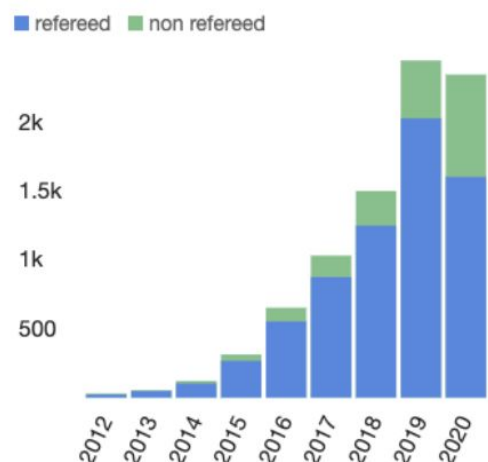

Fig 2. Number of papers in ADS citing software.

The PDS's own roadmap includes a recommendation in support of data citation and linking of data products with literature records in the ADS, which would then provide an enhanced level of data discovery for planetary data. Indexing data and software along with the literature that uses them allows scientists to discover their existence and connections with the relevant research. This allows, for instance, to find papers containing data sets that are hosted by different archives, contain data in different wavelengths, or have made use of a particular software package.

## Integrating Collections and Curation

One of the traditional tools of scholarship is the critically curated collection of facts relevant to a discipline.  Perhaps the most well known example is the Chemical Rubber Company's venerable Handbook of Chemistry and Physics. Typically these entities provide comprehensive, updated lists of relevant data on research objects of interest, such as molecules, stars, genes, seeds, fish, etc., and link the recorded measurements to the original sources. Some fields are large enough to support more than one overlapping (possibly commercial) service, such as SciFinder and Reaxys in chemistry. An example of a large, community-driven effort is the Particle Data Group in high energy physics.

While astronomers have long had a well-integrated information system which provides these data, Planetary Science does not have such a system. For decades astronomers have relied on the data products of the NASA/IPAC Extragalactic Database (NED) for galaxies, the SIMBAD and VizieR databases of the Strasbourg Data Center (CDS) for stars, galaxies and exoplanets, and more recently the NASA Exoplanet Archive at NExScI for exoplanets. As new observations for any of these objects are published, they are cross-identified or statistically associated with previous data and ingested into the appropriate knowledge base. These organizations are very well integrated into the information infrastructure of astronomy, routinely sharing data and methods with the journals, the ADS, and the archives.  This integration goes well beyond simple high-level linking of resources: an object query in the ADS, for example, triggers several real-time API queries to SIMBAD and NED. The ADS collates the responses and presents the user with a single combined view of search results federated across systems, along with a selection of indexed objects found in the returned list of results that can be used to further refine the user query (see figure 3), and to access the object's properties from SIMBAD and NED.

Fig. 3. The list of exoplanets found ADS search results is available via the integration of SIMBAD API searches in ADS

We suggest that the Planetary Science Community begin to establish a similar network of data centers, tasked with maintaining connections between Solar System Objects, their properties, observations, and the articles they appear in. This will greatly improve the discovery and retrieval of articles and measurements related to these objects, while providing the needed provenance of the data collected. These organizations would work together with the PDS, the ADS, the planetary science journals and other international organizations to enable and enhance Planetary research.

We suggest that as a first project a datacenter be founded to create and maintain a comprehensive database of measurements and literature references for the features in our solar system listed in the USGS/IAU Gazetteer of Planetary Nomenclature, and that the contents of this database be made available through an open API to ADS and other data centers. This would require that the Gazetteer be routinely updated and maintained, and that the ADS contain the full text of the relevant literature, and make it available for data mining. Having data links between this database and the PDS and other international archives would also improve the product.

# Recommendations

NASA has invested a substantial effort in collecting and archiving high-value Planetary data and the community it serves is committed to the FAIR data principles. The existence of curated high-level data products linked to the literature in the ADS will greatly increase their discoverability, re-use, and overall scientific impact at a fraction of the cost of the original missions and experiments, as multiple studies have shown for NASA Astrophysics Data (White et al, 2009; Henneken & Accomazzi, 2012; Rebull et al, 2019).

**Recommendation 1.** NASA should instruct the ADS to support the Planetary Science and Astrobiology literature at the same level of service that it supports the Astrophysics literature, with the goal of supporting cross-disciplinary scientific endeavors.

**Recommendation 2.** The public deposit and citation of data and software should be encouraged to follow the FAIR guiding principles.

**Recommendation 3.** To facilitate increased use and discovery, NASA should establish procedures to create and maintain linkages between Planetary Science archival data, software, and the literature which uses them. This should begin with a collaboration between the PDS nodes and the ADS.

**Recommendation 4.** NASA should take the lead in developing an international system of Planetary Science data centers, similar to the system in place for Astrophysics.

# References

Henneken, E.A. & Accomazzi, A. 2012. *Linking to Data: Effect on Citation Rates in Astronomy.* Astronomical Data Analysis Software and Systems XXI, 461, 763

Kurtz, M.J., Accomazzi, A., & Henneken, E.A. 2018. *Merging the Astrophysics and Planetary Science Information Systems.* arXiv:1803.03598

Muench, A., Accomazzi, A., Holm Nielsen, L., et al. 2020. *Asclepias: An Infrastructure Project to Improve Software Citation across Astronomy.* Astronomical Data Analysis Software and Systems XXVII, 522, 711

Rebull, L.M., Desai, V., Teplitz, H., et al. 2017. *NASA's Long-Term Astrophysics Data Archives.* arXiv:1709.09566

Tenopir, C., Rice, N.~M., Allard, S., et al. 2020. *Data sharing, management, use, and reuse: Practices and perceptions of scientists worldwide.* PLoS ONE, 15, e0229003

White, R.L., Accomazzi, A., Berriman, G.B., et al. 2009. *The High Impact of Astronomical Data Archives.* Astro2010: The Astronomy and Astrophysics Decadal Survey P64.

Wilkinson, M.D., Dumontier, M., Aalbersberg, I.J. et al. 2016. *The FAIR Guiding Principles for scientific data management and stewardship.* Scientific Data 3, 160018